\documentclass[aps,pra,twocolumn,groupedaddress,nofootinbib,longbibliography]{revtex4-1}
\usepackage{amsmath}
\usepackage{amssymb}
\usepackage{graphicx}
\usepackage{ntheorem}
\usepackage{mathtools}
\usepackage[T1]{fontenc}
\usepackage{physics}
\usepackage{subfig}
\usepackage{overpic}
\usepackage{epstopdf}
\usepackage{bm}
\usepackage{color,soul}
\usepackage[bookmarks=false]{hyperref}
\usepackage{xcolor}
\hypersetup{colorlinks=true,citecolor=blue,
linkcolor=blue,urlcolor=blue,pdfstartview=FitH,
bookmarksopen=true}

\begin{document}
\title{Entropic Uncertainty Relations with Quantum Memory in Accelerated Frames via Unruh--DeWitt Detectors}
\author{Ming-Ming Du$^1$}
\email{mingmingdu@njupt.edu.cn}
\author{Hong-Wei Li$^1$, Shu-Ting Shen$^1$, Xiao-Jing Yan$^2$, Xi-Yun Li$^2$, Lan Zhou$^2$,Wei Zhong$^3$}
\author{Yu-Bo Sheng$^{1,3}$}
\email{shengyb@njupt.edu.cn}
\affiliation{$1.$ College of Electronic and Optical Engineering and College of Flexible Electronics (Future Technology), Nanjing
University of Posts and Telecommunications, Nanjing, 210023, China\\
$2.$ School of Science, Nanjing University of Posts and Telecommunications, Nanjing,
210023, China\\
$3.$ Institute of Quantum Information and Technology, Nanjing University of Posts and Telecommunications, Nanjing, 210003, China}
\date{\today}
\begin{abstract}
Quantum uncertainty is deeply linked to quantum correlations and relativistic motion. The entropic uncertainty relation with quantum memory (QMA-EUR) offers a powerful way to study how shared entanglement affects measurement precision. However, under acceleration, the Unruh effect can degrade quantum correlations, raising questions about the reliability of QMA-EUR in such settings. Here, we investigate the QMA-EUR for two uniformly accelerating Unruh–DeWitt detectors coupled to a massless scalar field. Using the Kossakowski–Lindblad master equation, we calculate the entropic uncertainty, its lower bound, and the tightness of the relation under different Unruh temperatures. We find that acceleration does not always increase the lower bound on the uncertainty relation. Depending on the initial correlations between the detectors, it may either increase or decrease. This behavior results from the interplay between quantum discord and minimal missing information. Interestingly, a higher quantum discord does not necessarily lead to lower uncertainty.
\end{abstract}

\maketitle
\section{Introduction}
What if the vacuum of space is not truly empty? According to the Unruh effect, a uniformly accelerating observer perceives the vacuum as a thermal bath, experiencing particles in what would otherwise appear as empty space \cite{Unruh1976}. This remarkable prediction from relativistic quantum field theory reveals a deep link between acceleration, temperature, and quantum perception. Consider two observers uniformly accelerating through this vacuum, attempting to perform quantum measurements. How does their motion influence the fundamental quantum uncertainty associated with their measurements?

This question lies at the intersection of quantum information and relativistic physics. In particular, the entropic uncertainty relation (EUR) with quantum memory(QMA-EUR) \cite{Berta2010} offers a powerful framework for understanding how information, entanglement, and measurement precision interact in relativistic settings \cite{FuentesSchuller2005,Alsing2006,Pan2008,Datta2009,Landulfo2009,MartinMartinez2010,Chen2014,Crispino2008,Feng2013,Wu2022,Liu2025c,Tang2025,Li2025,Wu2024,Wu2024a,Wu2023,Liu2025,Liu2025a,Du2024,Du2024a,Li2024,Jia2015,Fan2019,Feng2015,Haseli2019,He2020,Wang2018,Huang2018,Qian2020,Shahbazi2020}. Prior studies have shown that quantum correlation can reduce measurement uncertainty \cite{Berta2010}. However, under relativistic motion, particularly in the presence of the Unruh effect, quantum correlations may degrade \cite{Landulfo2009,Crispino2008,Li2025,Zhou2021,Li2025a,Barman2022,Zhang2020,Benatti2004,Alsing2006,Pan2008}, and the predictive advantage of quantum memory may be compromised.

In this paper, we investigate the QMA-EUR for a pair of uniformly accelerating Unruh–DeWitt(UDW) detectors coupled to a massless scalar field in $(3+1)$-dimensional Minkowski spacetime \cite{Benatti2004}. We find that acceleration does not universally loosen the EUR bound. In some regimes, an increase in the Unruh temperature can tighten the uncertainty bound; in others, it loosens it—even as the uncertainty itself increases. To unravel this behavior, we analyze the contributions of quantum discord(QD) \cite{Ollivier2001,Henderson2001} and the minimal missing information(MMI) about the measured system. Our results reveal a subtle balance between these effects, highlighting that relativistic motion can both enhance and degrade uncertainty bounds in counterintuitive ways.

This study provides new insights into the interplay between relativistic effects and quantum information, suggesting that uncertainty, far from being merely a limitation, serves as a sensitive probe of relativistic quantum environments.

The remainder of this paper is organized as follows. In Sec.~\ref{sec:QMA}, we review the QMA-EUR. In Sec.~\ref{sec:model}, we introduce the UDW detector model. Sec.~\ref{sec:results} presents the numerical results for the entropic uncertainty, its lower bound, and the corresponding tightness under varying Unruh temperatures. Sec.~\ref{sec:discord}, we further analyze the roles of quantum discord and the minimum amount of missing information \( M \). Finally, Sec.~\ref{sec:conclusion} summarizes our findings.
\section{quantum‐memory‐assisted entropic uncertainty relation}\label{sec:QMA}
The uncertainty principle \cite{Heisenberg1927,Robertson1929} fundamentally distinguishes quantum from classical physics by asserting that certain pairs of observables cannot be simultaneously measured with arbitrary precision. Traditionally formulated in terms of standard deviations, this principle sets lower bounds on the product of measurement errors for noncommuting operators. While powerful, the standard deviation formulation has limitations when dealing with general probability distributions or information‐theoretic applications.

In the 1980s, Deutsch \cite{Deutsch1983} introduced an entropic perspective on quantum uncertainty, employing Shannon entropy to quantify the unpredictability of measurement outcomes. This approach was subsequently refined by Maassen and Uffink \cite{Maassen1988}, who derived an entropic uncertainty relation (EUR) of the form
$
H(X) + H(Z) \geq \log_2 \left(\frac{1}{c}\right),
$
where $H(\cdot)$ denotes the Shannon entropy of the measurement outcome distribution, and $c = \max_{i,j} |\langle x_i|z_j\rangle|^2$ quantifies the complementarity of the two measurement bases ${|x_i\rangle}$ and ${|z_j\rangle}$.  Note that some works instead define $c'=\max_{i,j}|\langle x_i | z_j \rangle|$ 
(without the square), in which case the MU bound reads 
$H(X)+H(Z)\ge -2 \log_2 c'$. 
Both conventions are equivalent. The EUR framework not only provides tighter, basis‐independent uncertainty bounds but also seamlessly integrates with classical and quantum information theory, underpinning protocols in quantum cryptography and randomness extraction \cite{Coles2017,Tomamichel2011}.

In 2010, Berta \emph{et al.} \cite{Berta2010} extended the entropic uncertainty relation to incorporate quantum side information. In their QMA–EUR, one considers a bipartite state $\rho_{AB}$ shared between a system $A$ (to be measured) and a memory $B$. Measurements $X$ and $Z$ performed on $A$ yield conditional entropies $S(X|B)$ and $S(Z|B)$, and the QMA–EUR takes the form
\begin{align}\label{qma-eur}
S(X|B) + S(Z|B) \geq \log_2 \left(\frac{1}{c}\right) + S(A|B)
\end{align}
where $S(X|B)=S(\rho_{XB})-S(\rho_B)$ and $S(Z|B)=S(\rho_{ZB})-S(\rho_B)$ are the conditional von-Neumann entropies of the post measurement states
\begin{align}\label{du1}
\rho_{XB} = \sum_{x_{i}} (|x_{i}\rangle\langle x_i| \otimes I) \rho_{AB} (|x_{i}\rangle\langle x_i| \otimes I),\\\notag
\rho_{ZB} = \sum_{z_{i}} (|z_{i}\rangle\langle z_i| \otimes I) \rho_{AB} (|z_{i}\rangle\langle z_i| \otimes I),
\end{align}
and
$S(A|B)=S(\rho_{AB})-S(\rho_B)$ is the conditional von Neumann entropy. Here and below, we denote the left-hand side of inequality Eq.~(\ref{qma-eur}) by $U$, and the right-hand side by $\mathcal{B}$. We define the tightness of the uncertainty relation as $\delta = U - \mathcal{B}$. If the memory $B$ is trivial and uncorrelated with $A$, then 
$S(X|B)=H(X)$, $S(Z|B)=H(Z)$, and $S(A|B)=S(A)$. 
For a pure state of $A$ (the state-independent MU setting) we have $S(A)=0$, 
and the inequality reduces to \(H(X)+H(Z) \ge \log_2 \tfrac{1}{c}\), which is exactly the Maassen--Uffink relation in the squared-overlap convention.

The presence of quantum memory can reduce the overall uncertainty, even allowing the sum of entropies to fall below the classical limit when $A$ and $B$ are entangled ($S(A|B)<0$). This relation has been experimentally tested \cite{Li2011} and applied in security proofs for quantum key distribution \cite{Berta2010,Tomamichel2011} and in witnessing entanglement \cite{Berta2010,Li2011,Prevedel2011}.

\section{Unruh–DeWitt detector model}\label{sec:model}
 We consider a system composed of two uniformly accelerating UDW detectors interacting with a massless scalar field in $(3+1)$-dimensional Minkowski spacetime \cite{Benatti2004}. Each detector is modeled as a two-level atom coupled to the quantum field, and the full system is treated as an open quantum system.

The total Hamiltonian of the system is given by,
\begin{equation}
	\mathcal{H} = \frac{\omega}{2} \Sigma_3 + \mathcal{H}_\Phi + \mu \mathcal{H}_I,
\end{equation}
where $\omega$ denotes the energy level spacing of each detector, and $\mu$ is the dimensionless coupling strength. $\Sigma_i$ is defined as
\begin{equation}
	\Sigma_i = \sigma_i^{(A)} \otimes I^{(B)} + I^{(A)} \otimes \sigma_i^{(B)},
\end{equation}
with $\sigma_i^{(\alpha)}$ ($\alpha = A,B$) being the Pauli matrices. $\mathcal{H}_\Phi$ represents the Hamiltonian of free massless scalar fields $\Phi(t,x)$ satisfying standard Klein-Gordon relativistic equation \cite{Benatti2004} and
\begin{equation}
	\mathcal{H}_I = (\sigma_2^{(A)} \otimes I^{(B)}) \Phi(x_A) + (I^{(A)} \otimes \sigma_2^{(B)}) \Phi(x_B)
\end{equation}
 is the interaction Hamiltonian. 
In this paper, we assume that the dimensionless coupling parameterder ($\mu\ll1$), indicating a weak coupling between the accelerated detectors and their surrounding environment.

 The initial state of the total system can be written as
\begin{equation}
	\rho_{\text{tot}}(0) = \rho_{AB}(0) \otimes |0\rangle \langle 0|,
\end{equation}
where $\rho_{AB}(0)$ is the initial state of the detectors, and $|0\rangle$ is the Minkowski vacuum state of the field. The evolution of the full system is governed by the von Neumann equation
\begin{equation}
	\dot{\rho}_{\text{tot}}(\tau) = -i [\mathcal{H}, \rho_{\text{tot}}(\tau)],
\end{equation}
where $\tau$ denotes the proper time along the detectors' trajectories. In the limit of weak coupling, the resulting evolution is Markovian and can be described by a Kossakowski–Lindblad master equation~\cite{Gorini1976,Lindblad1976}
\begin{equation}\label{mas}
	\frac{\partial \rho_{AB}(\tau)}{\partial \tau} = -i[\mathcal{H}_{\text{eff}}, \rho_{AB}(\tau)] + \mathcal{L}[\rho_{AB}(\tau)],
\end{equation}
with
\begin{equation}
	\mathcal{H}_{\text{eff}}=\frac{\omega}{2} \Sigma_3-\frac{i}{2}\Sigma_{ij}C_{ij}\sigma_i\sigma_j,
	\end{equation}
and
\begin{align}
	\mathcal{L}[\rho] = \sum_{i,j=1}^{3} \sum_{\alpha,\beta=A,B} \frac{C_{ij}}{2} \left( 2\sigma_j^{(\beta)} \rho \sigma_i^{(\alpha)} - \{\sigma_i^{(\alpha)} \sigma_j^{(\beta)}, \rho \} \right).
\end{align}
Here,  $C_{ij}$ is the element of Kossakowski matrix. These elements are determined by the Fourier transform of the Wightman function:
\begin{equation}
	G(\lambda) = \int_{-\infty}^{\infty} d\tau\, e^{i\lambda \tau} G^+(\tau), \quad G^+(\tau) = \langle 0 | \Phi(\tau) \Phi(0) | 0 \rangle.
\end{equation}
Using this Fourier transform, the matrix $C_{ij}$ can be decomposed as,
\begin{equation}\label{}
	C_{ij} = \frac{\gamma_+}{2}\delta_{ij} - i \frac{\gamma_-}{2}\epsilon_{ijk}\delta_{3k} + \gamma_0 \delta_{3i} \delta_{3j},
\end{equation}
with
\begin{equation}\label{du0}
	\gamma_\pm = G(\omega) \pm G(-\omega), \quad \gamma_0 = G(0) - \frac{\gamma_+}{2}.
\end{equation}

The effective Hamiltonian $\mathcal{H}_{\text{eff}}$ also contains a Lamb shift term:
\begin{equation}
	\mathcal{H}_{\text{eff}} = \frac{1}{2} \tilde{\omega} \sigma_3, \quad \tilde{\omega} = \omega + i[K(-\omega) - K(\omega)],
\end{equation}
where
\begin{equation}
	K(\lambda) = \frac{1}{i\pi} \mathcal{P} \int_{-\infty}^{\infty} d\omega \frac{G(\omega)}{\omega - \lambda},
\end{equation}
is the Hilbert transform of $G(\lambda)$.
Due to the thermal nature of the vacuum observed by an accelerating detector, the Wightman function satisfies the Kubo-Martin-Schwinger (KMS) condition, i.e.,
\begin{equation}
	G^+(\tau) = G^+(\tau + i\beta), \quad \beta = \frac{2\pi}{a} = \frac{1}{T},
\end{equation}
where $T$ is the Unruh temperature. Note that in our analysis we adopt natural units $\hbar = c = k = 1$, in which case the Unruh temperature $T$ is a dimensionless parameter. Translating it into frequency space,one has
\begin{equation}
	G(\lambda) = e^{\beta \lambda} G(-\lambda).
\end{equation}
 Using translation in-variance,  after some algebras, the Eq. (\ref{du0}) can be obtained as
\begin{align}
	\gamma_+ &= \left(1 + e^{-\beta \omega} \right) G(\omega), \\\notag
	\gamma_- &= \left(1 - e^{-\beta \omega} \right) G(\omega).
\end{align}	
For later use,we also introduce the ratio
\begin{equation}
\gamma= \frac{\gamma_-}{\gamma_+} = \tanh\left( \frac{\beta \omega}{2} \right).
\end{equation}

To obtain the stationary solution, we expand the two-qubit density operator on the Pauli tensor basis,
\begin{align}\label{r_ab}
\rho_{AB}(\tau)=&\tfrac14\bigg[I^{(A)}\otimes I^{(B)}+\sum_{i=1}^{3}\rho_{i0} \sigma^{(A)}_i\otimes I^{(B)}\\\notag
&+\sum_{i=1}^{3}\rho_{0i} I^{(A)}\otimes \sigma^{(B)}_i
+\sum_{i,j=1}^{3}\rho_{ij}\sigma^{(A)}_i\otimes\sigma^{(B)}_j\bigg],
\end{align}
where \(\rho_{i0}=tr[\rho_{AB}(\tau)\sigma^{(A)}_i\otimes I^{(B)}]\), \(\rho_{0i}=tr[\rho_{AB}(\tau)I^{(A)}\otimes \sigma^{(B)}_i]\), and \(\rho_{ij}=tr[\rho_{AB}(\tau)\sigma^{(A)}_i\otimes\sigma^{(B)}_j]\). Because the two detectors are identical and symmetrically coupled, the master equation preserves exchange symmetry ($A \leftrightarrow B$) as well as rotational invariance around the acceleration axis. As a result, the dynamics closes on the subspace
\begin{align}\label{du123}
\rho_{30}=\rho_{03}=u,~~ \rho_{11}=\rho_{22}=w,~~\rho_{33}=v,
\end{align}
with all other components vanishing. 
According to Eq. (\ref{mas}) and (\ref{du123}) yields three algebraic conditions:
\begin{align}
   &(3+\gamma^2)u+(3+\Delta_0)\gamma=0, \\\notag
   &(3+\gamma^2)w-(\Delta_0-\gamma^2)=0, \\\notag
   &(3+\gamma^2)v-[\Delta_0+(\Delta_0+2)\gamma^2]=0.
\end{align}
From these we obtain
\begin{align}
   u&=-\frac{(3+\Delta_0)\gamma}{3+\gamma^2},\\\notag
   w&=\frac{\Delta_0-\gamma^2}{3+\gamma^2},\\\notag
   v&=\frac{\Delta_0+(\Delta_0+2)\gamma^2}{3+\gamma^2}.
\end{align}
Here the parameter $\Delta_0$ encodes the initial two-qubit correlations and is defined by
$
\Delta_0=tr[\rho_{AB}(0)\,S], 
$
where $
S = \sigma^{(A)}_1 \otimes \sigma^{(B)}_1 + \sigma^{(A)}_2 \otimes \sigma^{(B)}_2 + \sigma^{(A)}_3 \otimes \sigma^{(B)}_3$. Using the identity $S=2V-I^{(A)}\otimes I^{(B)}$, where $V$ is the swap operator, one finds that the eigenvalues of $S$ are $\{1,-3\}$. 
Therefore, for any physical initial state $\rho_{AB}(0)$, the parameter satisfies the tight bounds
$
-3 \;\leq\; \Delta_0 \;\leq\; 1 .
$
The lower bound is attained by the singlet state $|\Psi^-\rangle$, while the upper bound corresponds to any triplet state.
Reconstructing the density matrix in the computational basis then gives the X-shaped stationary state
\begin{equation}\label{rab}
\rho_{AB}=
\begin{pmatrix}
x & 0 & 0 & 0 \\
0 & y & d & 0 \\
0 & d & y & 0 \\
0 & 0 & 0 & z
\end{pmatrix},
\end{equation}
with coefficients~\cite{Benatti2004}
\begin{align}
x &= \frac{(3 + \Delta_0)(\gamma - 1)^2}{4(3 + \gamma^2)}, \qquad
y = \frac{3 - \Delta_0 - (\Delta_0 + 1)\gamma^2}{4(3 + \gamma^2)}, \notag\\
z &= \frac{(3 + \Delta_0)(\gamma + 1)^2}{4(3 + \gamma^2)},\qquad
d = \frac{\Delta_0 - \gamma^2}{2(3 + \gamma^2)}.
\end{align}

\section{Numerical Results and Analysis}\label{sec:results}
To evaluate \( S(X|B) \) and \( S(Z|B) \), we perform projective measurements on qubit \( A \) in the eigenbases of \( \sigma_x \) and \( \sigma_z \), respectively. According to Eqs.~(\ref{du1}) and~(\ref{rab}), the conditional von Neumann entropy 
following a projective measurement in basis $M \in \{X,Z\}$ is given by
\begin{equation}
S(M|B) = H(p_m^M) + \sum_m p_m^M S(\rho_{MB|m}) - S(\rho_B),
\end{equation}
where $p_m^M$ are the measurement probabilities and 
$\rho_{MB|m}$ the corresponding post-measurement states. 
Explicitly, the uncertainty reads
\begin{align}\label{u}
U &= H(p_+,p_-) + p_+ S(\rho^{(+)}_B) + p_- S(\rho^{(-)}_B)  \notag \\
  &\quad + H(p_0,p_1) + p_0 S(\rho^{(0)}_B) + p_1 S(\rho^{(1)}_B) 
  - 2S(\rho_B),
\end{align}
with the Shannon entropies
\begin{align}
H(p_+,p_-) &= -p_+ \log_2 p_+ - p_- \log_2 p_-, \\\notag
H(p_0,p_1) &= -p_0 \log_2 p_0 - p_1 \log_2 p_1,
\end{align}
and the conditional entropies
\begin{align}\label{du2}
S(\rho^{(+)}_B) &= - \frac{x}{p_+} \log_2 \!\left(\frac{x}{p_+}\right) 
                  - \frac{y}{p_+} \log_2 \!\left(\frac{y}{p_+}\right), \\\notag
S(\rho^{(-)}_B) &= - \frac{y}{p_-} \log_2 \!\left(\frac{y}{p_-}\right) 
                  - \frac{z}{p_-} \log_2 \!\left(\frac{z}{p_-}\right), \\\notag
S(\rho^{(0)}_B) &= - \frac{x}{p_0} \log_2 \!\left(\frac{x}{p_0}\right) 
                  - \frac{y}{p_0} \log_2 \!\left(\frac{y}{p_0}\right), \\\notag
S(\rho^{(1)}_B) &= - \frac{y}{p_1} \log_2 \!\left(\frac{y}{p_1}\right) 
                  - \frac{z}{p_1} \log_2 \!\left(\frac{z}{p_1}\right),\\\notag
S(\rho_B) &= - \lambda_+^{(B)} \log_2 \lambda_+^{(B)} - \lambda_-^{(B)} \log_2 \lambda_-^{(B)}.
\end{align}
%
Here, the probabilities are
\begin{align}
p_+ &= \tfrac{1}{2}(x + 2y + z), \quad
p_- = \tfrac{1}{2}(1 - x - 2y - z), \notag\\
p_0 &= x + y, \quad
p_1 = y + z,
\end{align}
and the eigenvalues of \( \rho_B \) are
\begin{align}
\lambda_{\pm}^{(B)} = \tfrac{1}{2} \left[ 1 \pm \sqrt{(x - y)^2 + 4d^2} \right].
\end{align}

The entropy $S(\rho_{AB})$ of the bipartite system can be given as
\begin{align}\label{srab}
S(\rho_{AB}) =& -x \log_2 x - y \log_2 y - (z + d) \log_2 (z + d)\\\notag
&- (z - d) \log_2 (z - d).
\end{align}
According to Eq. (\ref{du2}) and Eq. (\ref{srab}), the conditional entropy $S(A|B)$ can be obtained as
\begin{align}\label{sraba}
S(A|B) = S(\rho_{AB}) - S(\rho_B).
\end{align}
Since $X$ and $Z$ correspond to Pauli operators with mutually unbiased bases, we have $c = \max_{i,j} |\langle x_i | z_j \rangle|^2 = \frac{1}{2}$.
Therefore, we have
\begin{align}\label{b}
\mathcal{B} = 1 + S(A|B).
\end{align}
According to Eq. (\ref{u}) and (\ref{b}), we can obtain the tightness as
\begin{align}
\delta=U-\mathcal{B}.
\end{align}

\begin{figure*}[t]
\subfloat[$\Delta_0=-1$]{\includegraphics[width=0.32\linewidth]{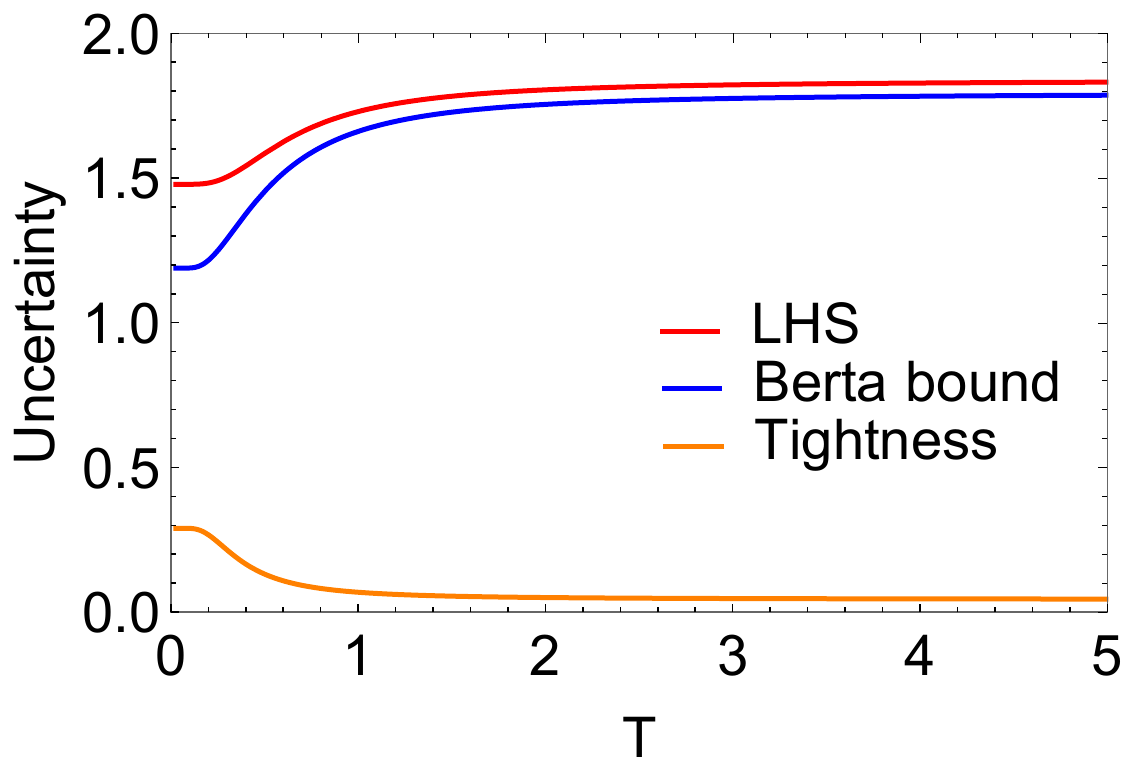}}
\subfloat[$\Delta_0=0.5$]{\includegraphics[width=0.32\linewidth]{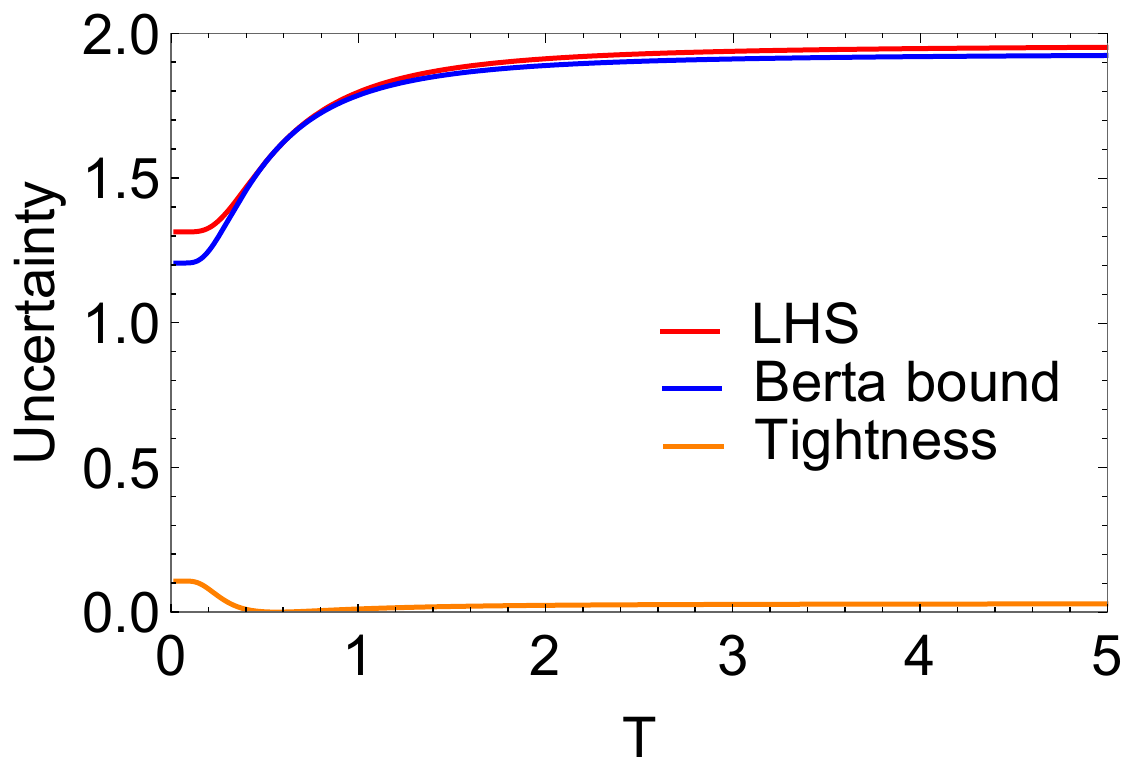}}
\subfloat[$\Delta_0=1$]{\includegraphics[width=0.32\linewidth]{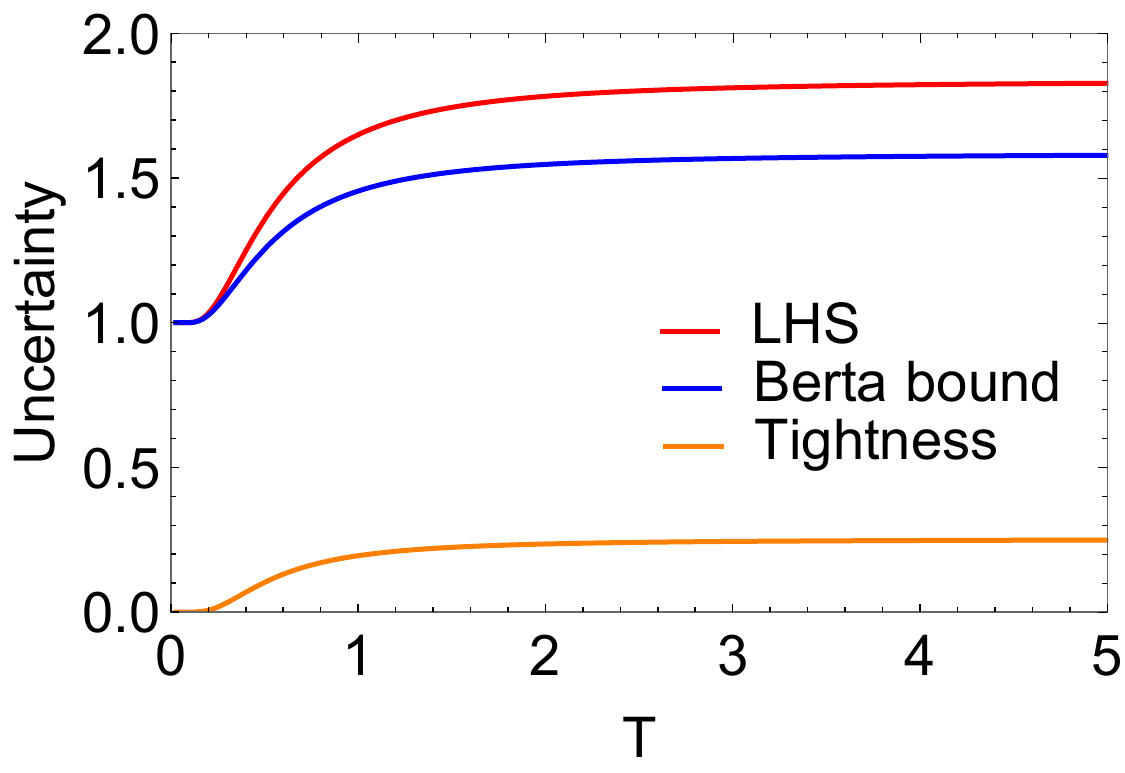}}
\caption{The uncertainty $U$, its bound $\mathcal{B}$ and its tightness $\delta$ as a function of the Unruh temperature $T$, for different values of the initial state selection parameter: (a)$\Delta_0 =-1$; (b)$\Delta_0 =0.5$; (c)$\Delta_0 =1$. In all numerics we set $\omega=1$.} 
\label{fig1}
\end{figure*}

\begin{figure*}[t]
	\subfloat[$\Delta_0=-1$]{\includegraphics[width=0.32\linewidth]{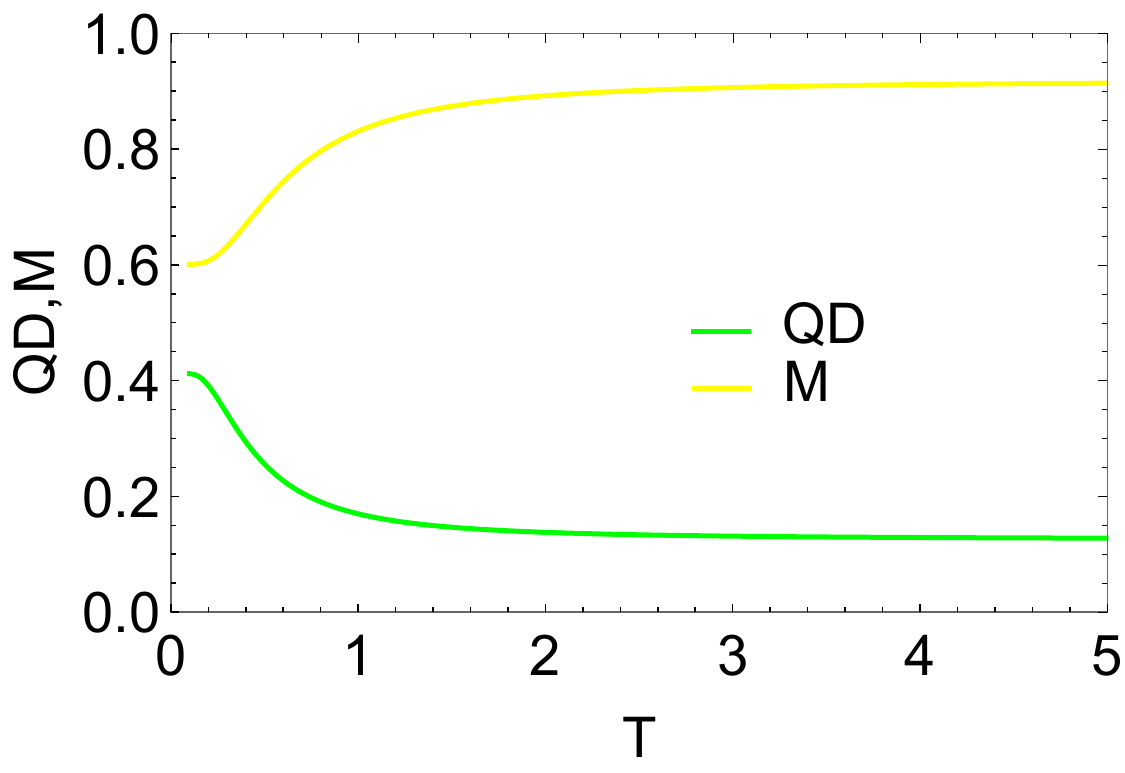}}
	\subfloat[$\Delta_0=0.5$]{\includegraphics[width=0.32\linewidth]{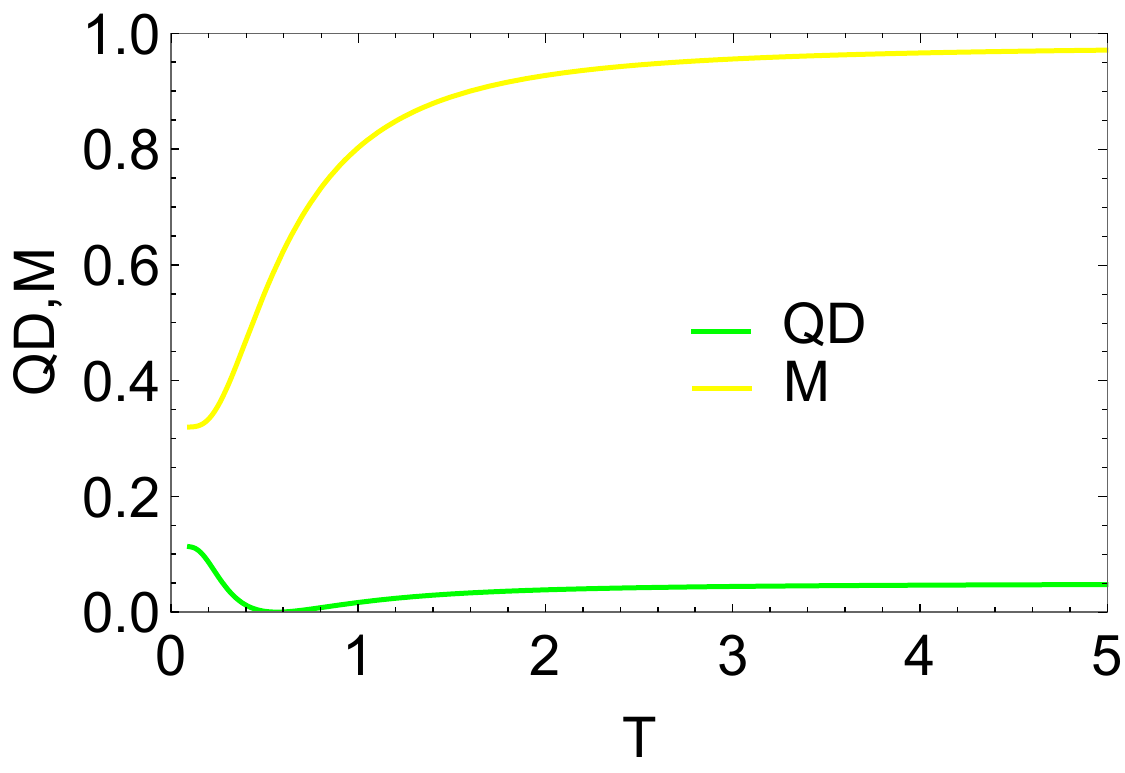}}
	\subfloat[$\Delta_0=1$]{\includegraphics[width=0.32\linewidth]{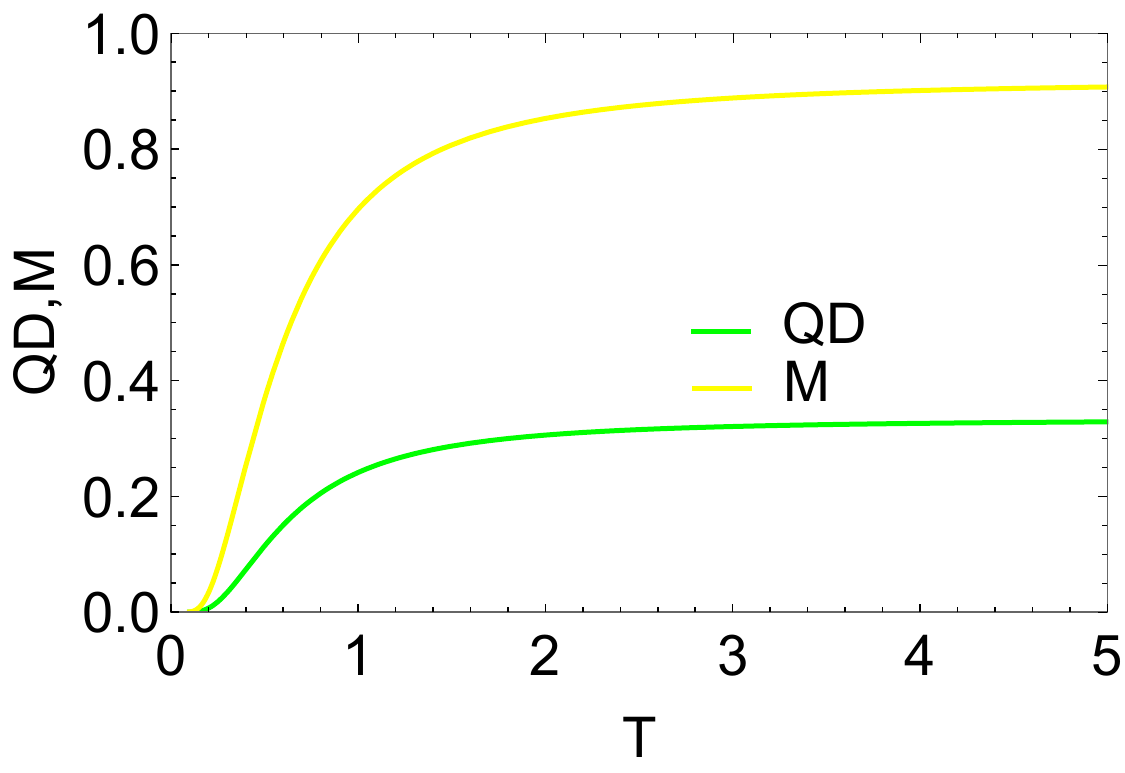}}
	\caption{The dynamics of the quantum discord and the minimal missing information as a function of the Unruh temperature $T$, for different values of the initial state parameter: (a)$\Delta_0 =-1$; (b)$\Delta_0 =0.5$; (c)$\Delta_0 =1$. In all numerics we set $\omega=1$.}
	\label{fig2}
\end{figure*}

To examine the impact of the Unruh effect on the QMA-EUR, we numerically evaluate the entropy uncertainty $U$, its lower bound $\mathcal{B}$, and the tightness $\delta = U - \mathcal{B}$ as functions of the Unruh temperature $T$. The results are presented in Fig. \ref{fig1}, where each panel corresponds to a different value of the initial state parameter $\Delta_0$. In Fig. \ref{fig1}(a), we fix $\Delta_0 = -1$. The entropy uncertainty increases monotonically with $T$. At low temperatures, the inequality is far from saturation, with a significant gap between $U$ and $\mathcal{B}$; however, as $T$ increases, this gap gradually decreases, indicating improved tightness. Fig. \ref{fig1}(b) displays the case $\Delta_0 = 0.5$. Here, $U$ again increases with $T$. The uncertainty relation initially becomes saturated as $T$ rises, but turns unsaturated again at higher temperatures, suggesting a nontrivial dependence of tightness on $T$. In Fig. \ref{fig1}(c), for $\Delta_0 = 1$, the uncertainty relation is initially saturated at $T = 0$, but becomes progressively less tight as $T$ increases, even though $U$ continues to grow. This indicates that the Unruh effect may degrade the effectiveness of the lower bound in certain regimes.

\section{Role of quantum discord and missing information in the entropic uncertainty relation}\label{sec:discord}
To better understand the behavior of the quantum-memory-assisted entropic uncertainty relation, 
it is useful to introduce the concepts of quantum discord (QD) and the minimum missing information $M$. 

For a bipartite state $\rho_{AB}$, the total correlations are captured by the quantum mutual information
\begin{align}
I(\rho_{AB}) = S(\rho_A) + S(\rho_B) - S(\rho_{AB}).
\end{align}
The classical correlations are defined as
\begin{align}
J(\rho_{AB}) = S(\rho_B) - \min_{\{B_k\}} \sum_k q_k S(\rho_A^k),
\end{align}
where $\{B_k\}$ is a set of projective measurements on subsystem $B$, 
$q_k = \mathrm{Tr}_{AB}(B_k \rho_{AB} B_k^\dagger)$ is the probability of outcome $k$, 
and $\rho_A^k = \mathrm{Tr}_B(B_k \rho_{AB} B_k^\dagger)/q_k$ is the corresponding post-measurement state. 

The quantum discord is then defined as
\begin{align}
D(\rho_{AB}) &= I(\rho_{AB}) - J(\rho_{AB}) \\\notag
&= -S(A|B) + \min_{\{B_k\}} \sum_k q_k S(\rho_A^k).
\end{align}
Here, the second term
\begin{align}
M = \min_{\{B_k\}} \sum_k q_k S(\rho_A^k)
\end{align}
quantifies the minimum amount of missing information about subsystem $A$ after a measurement on $B$. 

With these definitions, the lower bound of the QMA–EUR can be rewritten in the compact form
\begin{align}\label{bdm}
\mathcal{B} = \log_2 \frac{1}{c} + M - D.
\end{align}
This equivalent expression shows that the uncertainty bound is determined by the competition 
between the growth of missing information $M$ and the change in quantum discord $D$.

To better understand the behavior of uncertainty, we further analyze the roles of quantum discord and the minimum amount of missing information \( M \), as shown in Fig.~\ref{fig2}. Each panel corresponds to a different initial value of \( \Delta_0 \), consistent with Fig.~\ref{fig1}. In Fig.~\ref{fig2}(a), quantum discord decreases monotonically with increasing \( T \), while \( M \) increases. This leads to a consistent growth in the uncertainty, reflecting the loss of quantum correlations and the rise in inaccessible information due to the Unruh effect. In Fig.~\ref{fig2}(b), the quantum discord exhibits non-monotonic behavior—it first decreases and then increases—while \( M \) increases steadily. This results in a more complex uncertainty profile, governed by the competing effects of correlation recovery and information degradation. In contrast, Fig.~\ref{fig2}(c) shows that both quantum discord and \( M \) increase with \( T \), yet the rising missing information may outweigh the effect of increased correlations, leading to overall growth in uncertainty.  

These results clearly demonstrate that the entropic uncertainty is not solely dictated by the strength of quantum correlations. Specifically, an increase (or decrease) in quantum discord does not necessarily imply a corresponding decrease (or increase) in uncertainty. Rather, the uncertainty behavior is determined by the competition between quantum correlations and the degree of missing information induced by the Unruh effect. This is fully consistent with the analytical expression of the lower bound $\mathcal{B} = \log_2 (1/c) + M - D$ introduced in Eq. (\ref{bdm}), which explicitly reveals the interplay between discord and missing information.  

\section{Conclusion}\label{sec:conclusion}
In this paper, we have investigated the QMA-EUR for two uniformly accelerating UDW detectors interacting with a massless scalar field. We have numerically evaluated the entropic uncertainty, its lower bound, and the tightness of the relation as functions of the Unruh temperature $T$, across various initial correlation parameters $\Delta_0$. Our results reveal that the Unruh effect generally increases the uncertainty, but its influence on the tightness of the QMA-EUR is non-monotonic and highly sensitive to the initial detector correlations. Specifically, we identified regimes where the bound becomes tighter with increasing temperature, and others where it becomes looser, even as the overall uncertainty increases.

To deepen our understanding, we analyzed the roles of QD and the MMI. We found that the behavior of the uncertainty is governed by a competition between the decay or revival of quantum correlations and the increase in MMI induced by the Unruh effect. Interestingly, an increase in quantum discord does not always correspond to a reduction in uncertainty, nor does a decrease necessarily lead to its enhancement. This highlights the subtle and intricate interplay between quantum information measures and relativistic motion.

Finally, we remark on the use of the Markovian approximation in our analysis. Throughout this work we have assumed weak coupling between the detectors and the field, which allows the dynamics to be described by a Markovian Kossakowski–Lindblad master equation and makes analytic progress possible. Nevertheless, at very large accelerations the Unruh effect introduces a natural time scale, and the field correlation time may become comparable to the detector dynamics, potentially invalidating the Markovian assumption. In such regimes, non-Markovian approaches would be required for a more accurate description. Extending the present framework to include these effects is an important direction for future investigation.
\begin{acknowledgments}
This work was supported by the National Natural Science Foundation of China(Grant Nos. 12175106 and 92365110), the Natural Science Foundation of Jiangsu Province, China(Grant No.BK20240612), the Natural Science Research Start-up Foundation of Recruiting Talents of Nanjing University of Posts and Telecommunications(Grant No. NY222123), and the Natural Science Foundation of Nanjing University of Posts and Telecommunications(Grant No. NY223069).
\end{acknowledgments}
%
\end{document}